\documentclass[12pt]{article}
\usepackage{graphicx}
\usepackage{xspace}
\usepackage[numbers]{natbib}
\newcommand\pubnumber{NuPhys2018-Dealtry}
\newcommand\pubdate{\today}

\def\lancaster{Physics  Department\\Lancaster  University,  Lancaster,  United  Kingdom}

\def\Title#1{\begin{center} {\Large #1 } \end{center}}
\def\Author#1{\begin{center}{ \sc #1} \end{center}}
\def\Address#1{\begin{center}{ \it #1} \end{center}}

\newcommand\pubblock{\rightline{\begin{tabular}{l} \pubnumber\\
         \pubdate  \end{tabular}}}
\newenvironment{Abstract}{\begin{quotation}  }{\end{quotation}}
\newenvironment{Presented}{\begin{quotation} \begin{center} 
             PRESENTED AT\end{center}\bigskip 
      \begin{center}\begin{large}}{\end{large}\end{center} \end{quotation}}

\def\beq{\begin{equation}}
\def\eeq#1{\label{#1}\end{equation}}
\def\eeqn{\end{equation}}

\def\beqa{\begin{eqnarray}}
\def\eeqa#1{\label{#1}\end{eqnarray}}
\def\eeqan{\end{eqnarray}}

\let\bar=\overbar

\def\Dslash{\not{\hbox{\kern-4pt $D$}}}
\def\dslash{\not{\hbox{\kern-2pt $\del$}}}

\def\msb{{\bar{\ssstyle M \kern -1pt S}}}

\def\hk{Hyper-K\xspace}

\def\elike{$e$-like\xspace}
\def\mulike{$\mu$-like\xspace}

\newcommand{\dcp}     {\ensuremath{\delta_{CP}}\xspace}
\def\sinsqTthetasolar       {\ensuremath{\textrm{sin}^22\theta_{12}}\xspace}
\def\sinsqTthetareactor     {\ensuremath{\textrm{sin}^22\theta_{13}}\xspace} 
 
\def\sinsqthetaatmos        {\ensuremath{\textrm{sin}^2\theta_{23}}\xspace} 
\def\dmsqsolar     {\ensuremath{\Delta m^2_{21}}\xspace} 
\def\dmsqatmos     {\ensuremath{\Delta m^{2}_{32}}\xspace}

\def\nub        {\ensuremath{\overline{\nu}}\xspace}
\def\nubar      {\nub}
\def\nue        {\ensuremath{\nu_e}\xspace}
\def\nueb       {\ensuremath{\nub_e}\xspace}
\def\nuebar     {\nueb}
\def\num        {\ensuremath{\nu_\mu}\xspace}
\def\numu       {\num}
\def\numb       {\ensuremath{\nub_\mu}\xspace}
\def\numubar    {\numb}

\begin{document}
\begin{titlepage}
\pubblock

\vfill
\Title{Hyper-Kamiokande}
\vfill
\Author{T. Dealtry, for the Hyper-Kamiokande experiment}
\Address{\lancaster}
\vfill
\begin{Abstract}
  Hyper-Kamiokande (\hk) is a next generation large water Cherenkov detector  to  be  built  in  Japan,  based  on  the  highly  successful  Super-Kamiokande detector.  \hk will offer a broad science program including neutrino oscillation studies, proton decay searches, and neutrino astrophysics with unprecedented sensitivities.  This paper provides some highlights of the physics potential of \hk.
\end{Abstract}
\vfill
\begin{Presented}
  NuPhys2018, Prospects in Neutrino Physics
  Cavendish Conference Centre, London, UK, December 19--21, 2018
\end{Presented}
\vfill
\end{titlepage}
\def\thefootnote{\fnsymbol{footnote}}
\setcounter{footnote}{0}

\section{Introduction}

Hyper-Kamiokande (\hk) is a next generation water Cherenkov detector.
It has multiple improvements with respect to Super-K, the most important being an increase in fiducial volume of more than 8 times (187\,kton), and photomultiplier tubes (PMTs) single-photon detection efficiency improved by a factor two.
\hk construction will commence in April 2020, and data-taking is expected from $\sim$2027.

In these proceedings we present some highlights of the physics potential of the experiment. More details can be found in the Hyper-Kamiokande design report \cite{hkdr}.

\section{Beam neutrino physics potential}

The \hk detector will be located in the Tochibora mine, near Kamioka, Japan.
At this location it samples the J-PARC neutrino beam, 2.5$^\circ$ off-axis at a distance of 295\,km.
On- and off-axis detectors will be used to characterise and constrain the unoscillated neutrino beam, including INGRID \cite{ingrid}, ND280 upgrade \cite{nd280up}, and IWCD \cite{e61}.
The J-PARC beam is currently undergoing staged upgrades, and is expected to reach a beam power of 1.3 MW by $\sim$2027.
Results are shown for $2.7 \times 10^{22}$ protons on target, corresponding to 10 years of operation, split 1:3 between $\nu$:\nubar running.

Fully-contained single-ring events with vertices within the fiducial volume are selected.
Particle identification is performed on the ring to determine whether the event is \elike or \mulike.
Neutrino energy is reconstructed assuming the event was charged-current quasi-elastic (CCQE).
More details can be found in \cite{t2k_fitqun}.

The expected event rates for \elike and \mulike samples are shown in Tables \ref{tab:beam_nue} and \ref{tab:beam_numu} respectively.
These numbers assume the oscillation parameter values: 
$\dcp = 0$;
$\sinsqTthetareactor = 0.1$;
$\sinsqthetaatmos = 0.5$;
$\dmsqatmos = 2.4 \times 10^{-3}$\,eV$^2$;
$\sinsqTthetasolar = 0.8704$;
$\dmsqsolar = 7.6\times10^{-5}$\,eV$^2$;
normal mass hierarchy.

\begin{figure}[h!]
  \centering
  \includegraphics[height=9.5in,keepaspectratio,width=0.49\textwidth, trim={0 4.25in 0 0.5in}, clip]{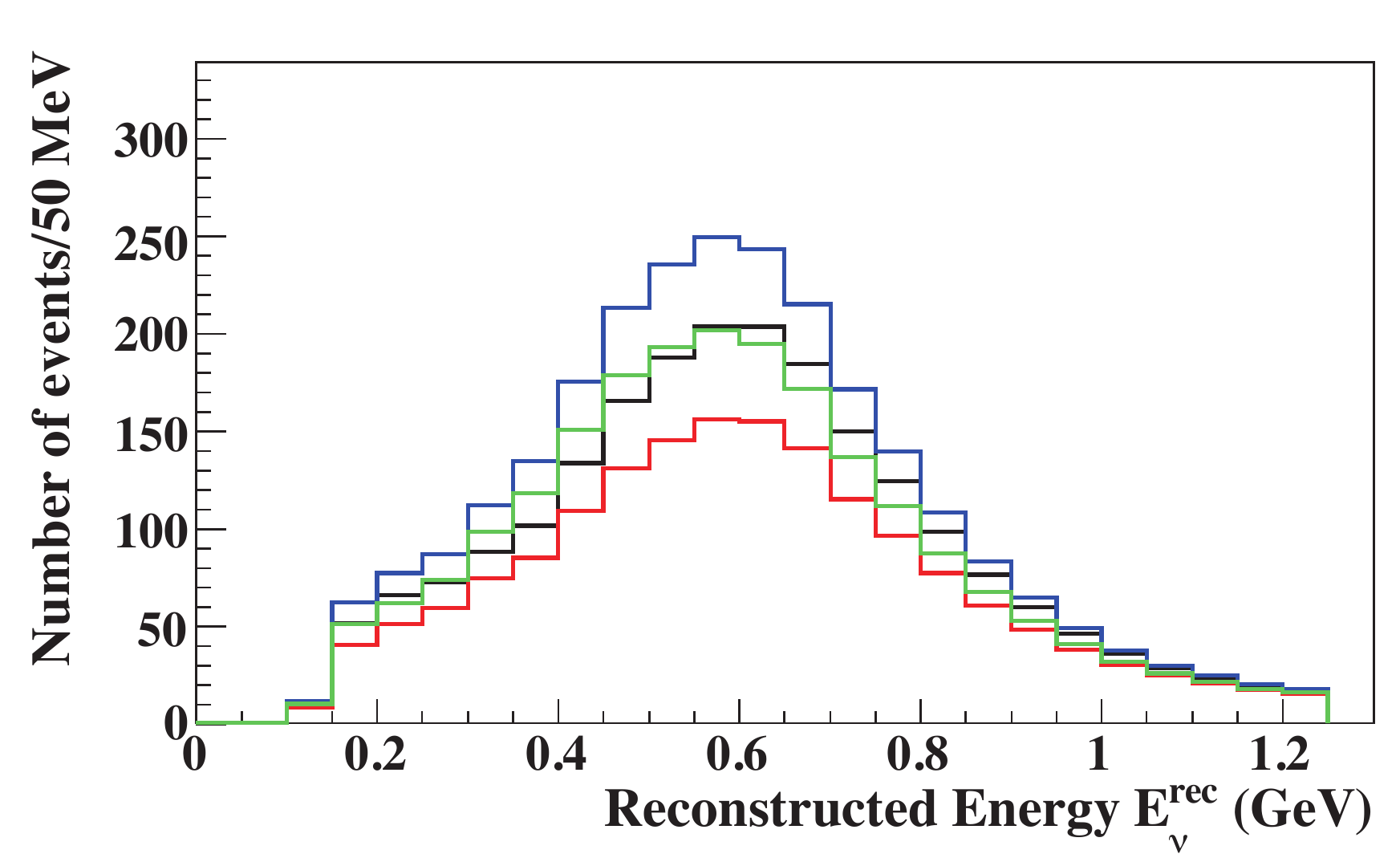}
  \includegraphics[height=9.5in,keepaspectratio,width=0.49\textwidth, trim={0 0 0 4.75in}, clip]{images/DeltaCP_Energy_numode_Appearence_1tank}
  \caption{
    Left: Reconstructed neutrino energy distribution for $\dcp = -90^\circ$, $0^\circ$, $90^\circ$, $180^\circ$ for the $\nu$ beam mode.
    Right: Difference to the $\dcp = 0^\circ$ case, with statistical uncertainties shown as error bars.
  }
  \label{fig:beam_nue_dcp}
\end{figure}

Figure \ref{fig:beam_nue_dcp} shows the effect of varying \dcp on the \elike event samples.
With statistical-only uncertainties, these extreme values of \dcp can be distinguished.

\begin{table}[b]
  \begin{center}
    \begin{tabular}{|c | c || c | c | c | c || c || c |}
      \hline
      & Right-sign & Wrong-sign & \numu/\numubar & Beam & NC & Total & T2K\\
      & $\numu \rightarrow \nue$ CC & $\numu \rightarrow \nue$ CC & CC & \nue/\nuebar CC & & & 2018 \\
      \hline
      $\nu$ beam & 1643 & 15 & 7 & 259 & 134 & 2058 & 75 \\
      \hline
      \nubar beam & 1183 & 206 & 4 & 317 & 196 & 1906 & 9 \\
      \hline
    \end{tabular}
    \caption{Expected number of \elike event candidates for each neutrino flavour and interaction type
      in the $\nu$ and \nub beams.
    T2K 2018 numbers are taken from \cite{t2k_neutrino2018}.}
    \label{tab:beam_nue}
  \end{center}
\end{table}

\begin{table}[htb]
  \begin{center}
    \begin{tabular}{|c || c | c | c | c | c | c || c || c |}
      \hline
      & Right-sign & Wrong-sign & \numu/\numubar & Beam & NC & $\numu \rightarrow$ & Total & T2K \\
      & CCQE & CCQE & CCnQE & \nue/\nuebar & & \nue CC & & 2018 \\
      \hline
      $\nu$ beam & 6043 & 348 & 3175 & 6 & 480 & 29 & 10080 & 243 \\
      \hline
      \hline
      \nubar beam & 6099 & 2699 & 4315 & 7 & 603 & 4 & 13726 & 102 \\
      \hline
    \end{tabular}
    \caption{Expected number of \mulike event candidates for each neutrino flavour and interaction type
      in the $\nu$ and \nub beams.
      T2K 2018 numbers are taken from \cite{t2k_neutrino2018}.}
    \label{tab:beam_numu}
  \end{center}
\end{table}



Systematic uncertainties are included in the analysis, based on experience from T2K \cite{t2k_sensi}.
The systematics are split into three categories, and assumptions are applied to each based on how improvements will be made with new data and analyses.

\emph{1) Flux and cross-section uncertainties constrained by a fit to near-detector data.} These are expected to be reduced as understanding is improved, however more uncertainties will be added from 2). The level is assumed to stay the same.

\emph{2) Cross-section uncertainties non-constrained by the near-detector fit.} New selections will be added to the near-detector fit, allowing movement from this category to 1). The level is assumed to be reduced.

\emph{3)Far detector efficiency and reconstruction uncertainties.} Current uncertainties are limited by atmospheric neutrino statistics, therefore in \hk these are assumed to reduce. The energy scale uncertainty is left unchanged.

The total uncertainty on the number of events is in the range 3.2--3.9\%, and is dominated by category 1).
\hk will be systematics limited, therefore it is important to have a realistic and comprehensive error model; work is underway to improve the above error model.

\begin{figure}[tb]
  \centering
  \includegraphics[height=9.5in,keepaspectratio,width=0.49\textwidth]{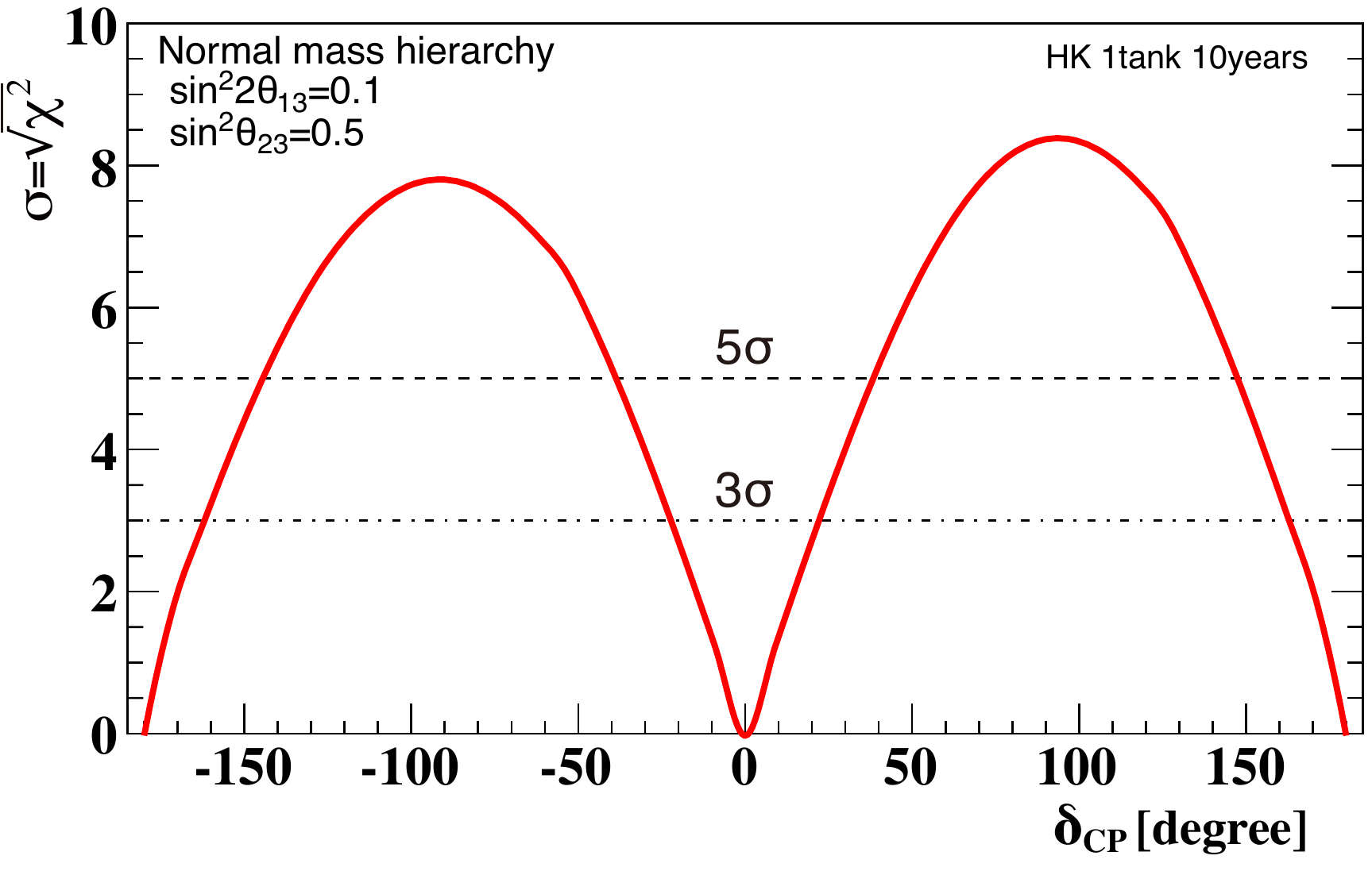}
  \includegraphics[height=9.5in,keepaspectratio,width=0.49\textwidth]{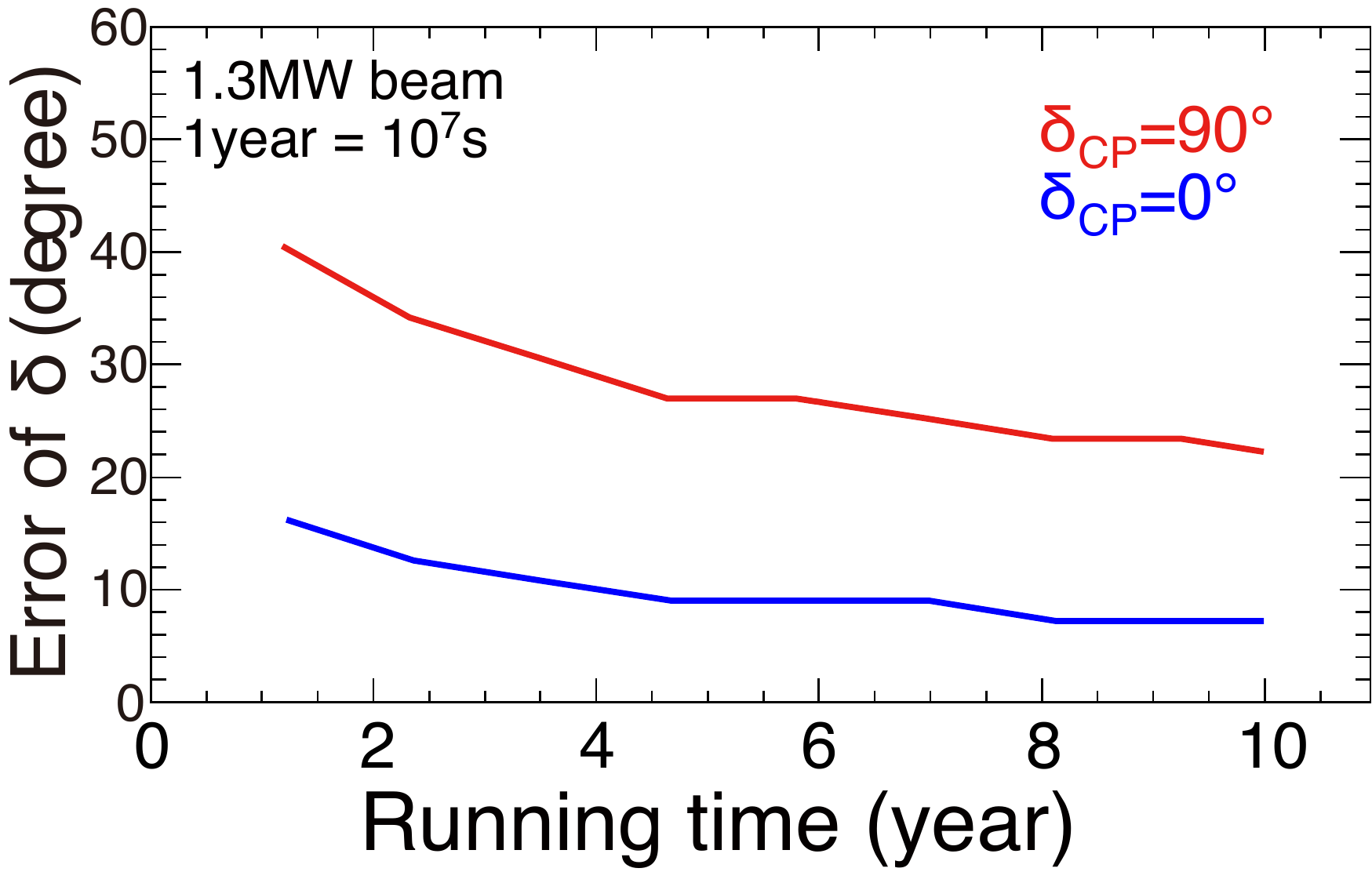}
  \caption{Left: Expected significance to exclude $\sin\dcp=0$, for all possible true values of \dcp.
    Right: Expected 1$\sigma$ error of \dcp as a fraction of running time.
    In both cases the mass hierarchy is assumed to be known and normal.}
  \label{fig:beam_dcp_sensi}
\end{figure}

The significance to exclude $\sin\dcp=0$ (i.e. CP conservation) is shown in the left hand side of Figure \ref{fig:beam_dcp_sensi}.
CP violation is observable with greater than 3(5)\,$\sigma$ significance for 76(57)\% of possible values of \dcp.
The uncertainty on \dcp is shown in the right hand side of Figure \ref{fig:beam_dcp_sensi}.
\dcp can be measured with an uncertainty of 7.2$^\circ$ at $\dcp=0^\circ$ and 23$^\circ$ at $\dcp=\pm90^\circ$.


\section{Atmospheric neutrino physics potential}

\begin{figure}[tb]
  \centering
  \includegraphics[height=9.5in,keepaspectratio,width=\textwidth]{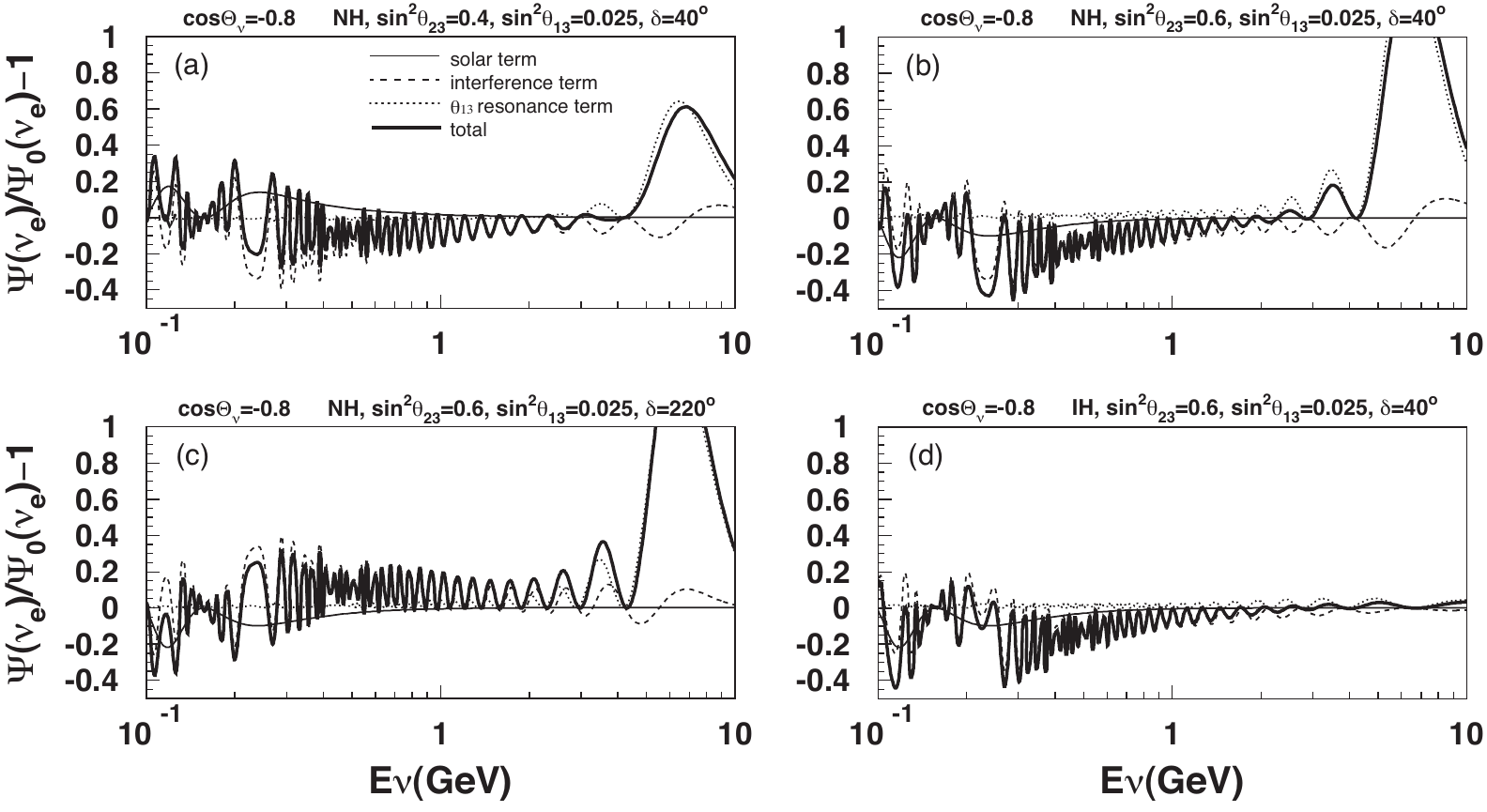}
  \caption{The ratio of oscillated to non-oscillated \nue flux for zenith angle $\cos\Theta_\nu = -0.8$ for various oscillation scenarios.
    The \nuebar flux is not included.}
  \label{fig:atmos_flux}
\end{figure}

The ratio of oscillated to non-oscillated \nue flux is shown in Figure \ref{fig:atmos_flux} for zenith angle $\cos\Theta_\nu = -0.8$ for various oscillation scenarios.
Sensitivity to the $\theta_{23}$ octant can be seen by comparing (b) $\sinsqthetaatmos = 0.6$ to (a) $\sinsqthetaatmos = 0.4$; the size of the high-energy resonance changes.
Sensitivity to \dcp can be seen by comparing (b) $\dcp = 40^\circ$ to (c) $\dcp = 220^\circ$; the scale and direction of $\sim$1\,GeV interference changes.
Sensitivity to the mass hierarchy can be seen by comparing (b) normal to (d) inverted; the high-energy resonance is only present in normal hierarchy \nue (and inverted hierarchy \nuebar).

\begin{figure}[htb]
  \centering
  \includegraphics[height=9.5in,keepaspectratio,width=\textwidth]{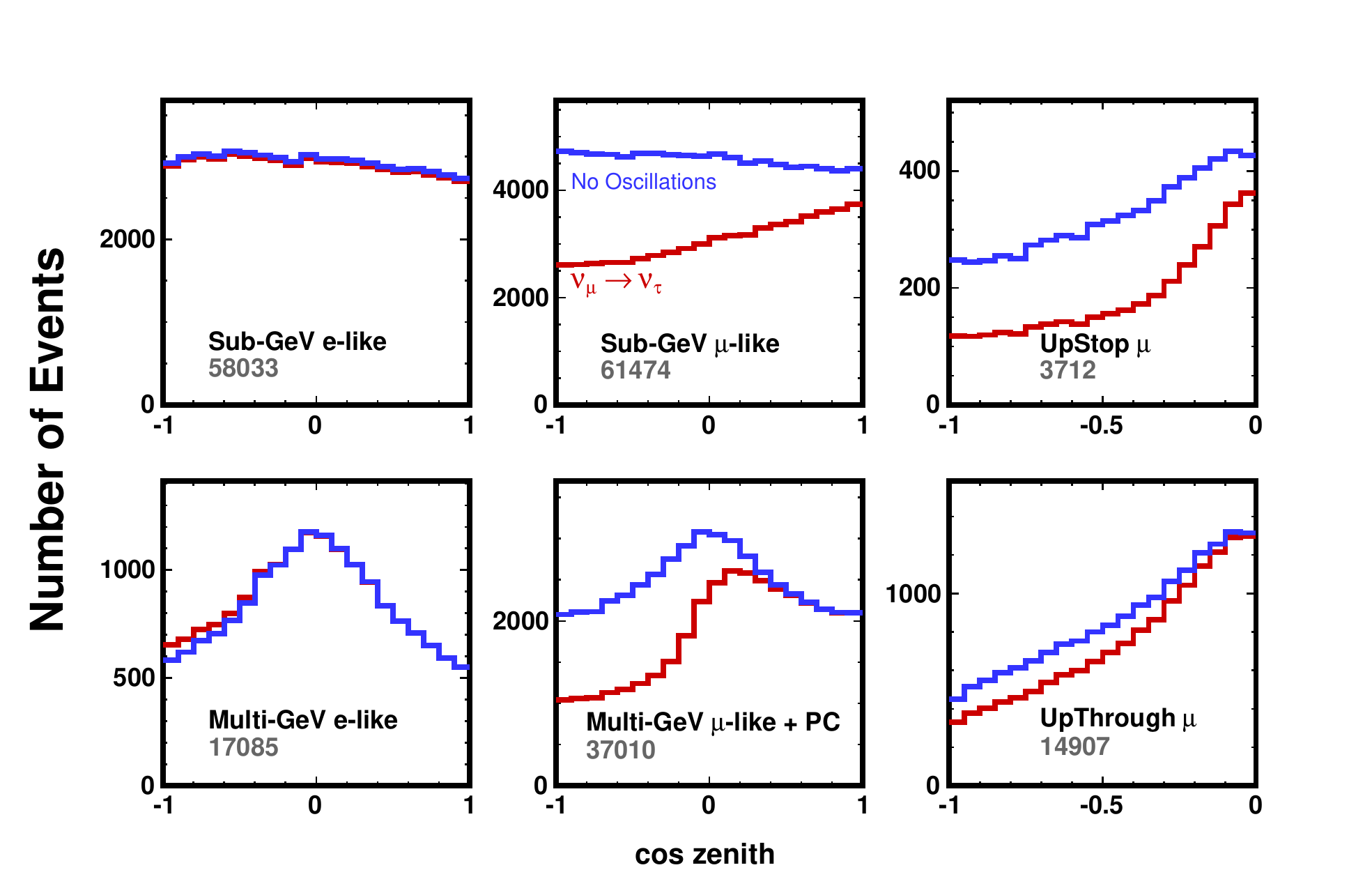}
  \caption{Zenith-angle distributions for 6 of 19 atmospheric neutrino analysis samples.
    Expectations with (blue) and without (red) oscillations are shown.
    Grey numbers indicate the integrated event rate of each sample.}
  \label{fig:atmos_samples}
\end{figure}

The atmospheric neutrino analysis selects 19 event samples using reconstructed neutrino energy, particle ID of the highest energy ring, the number of rings, and containment. More details can be found in \cite{sk_atmos}.
The zenith-angle distribution for 6 of the samples is shown in Figure \ref{fig:atmos_samples}.

\begin{figure}[h!]
  \centering
  \includegraphics[height=9.5in,keepaspectratio,width=0.49\textwidth]{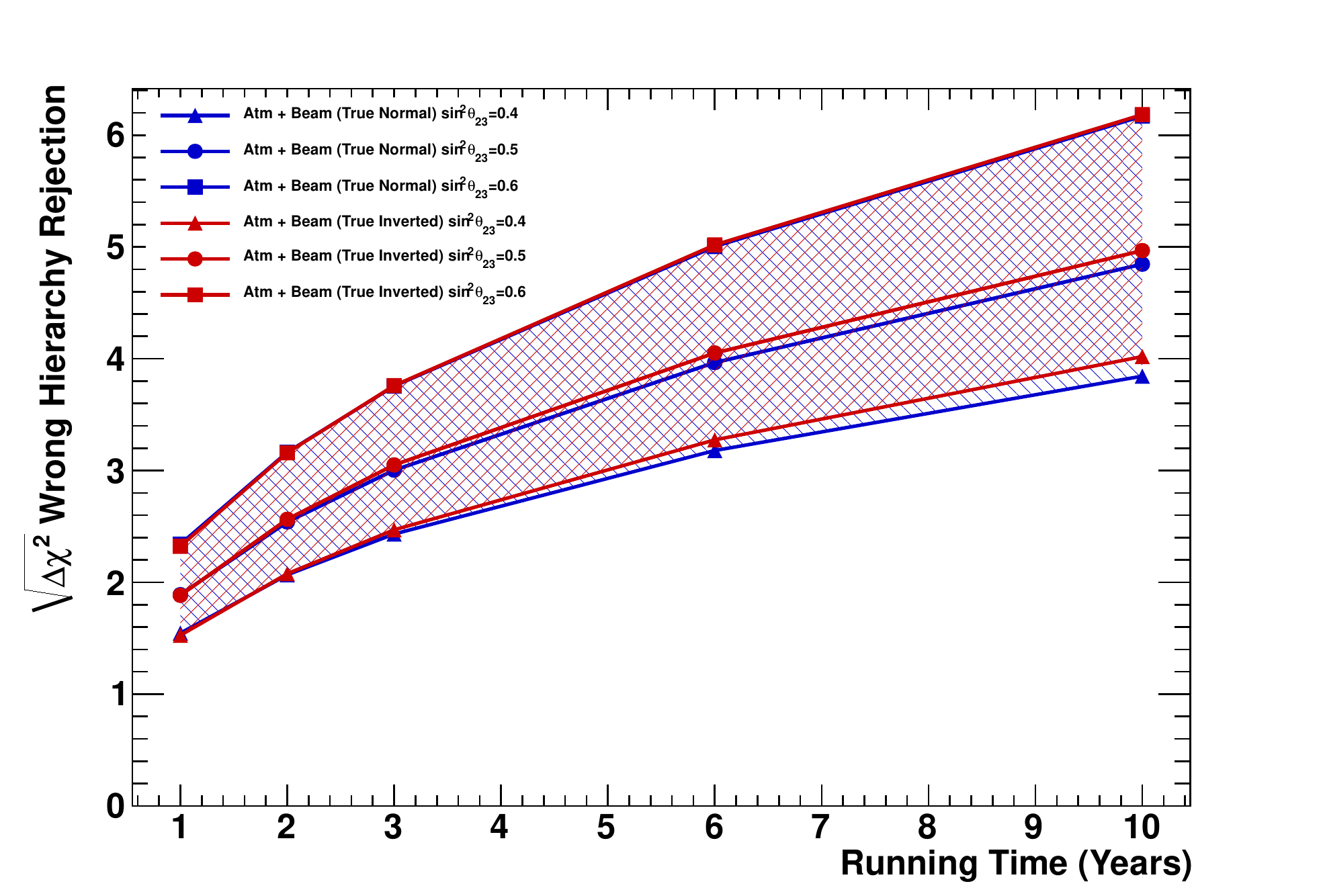}
  \includegraphics[height=9.5in,keepaspectratio,width=0.49\textwidth]{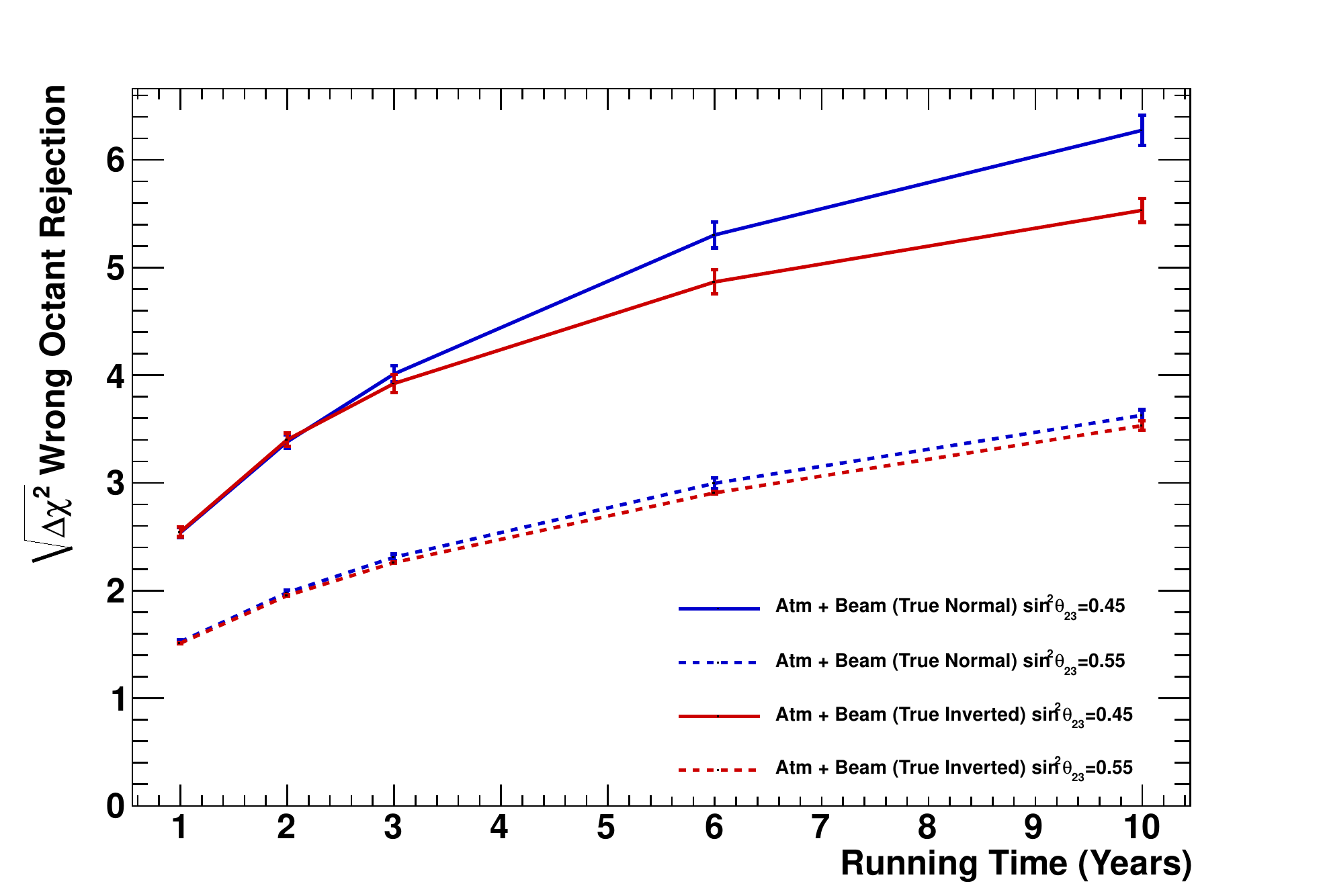}
  \caption{Left: Mass hierarchy sensitivity as a function of time for $\sinsqthetaatmos = 0.4$, $0.5$, and $0.6$.
    Right: Octant determination sensitivity as a function of time for $\sinsqthetaatmos = 0.45$ and $0.55$.
    In both plots, normal (inverted) hierarchy is shown in blue (red).}
  \label{fig:atmos_beam_sensi}
\end{figure}

The sensitivity to the oscillation parameters of interest is greatest when beam and atmospheric neutrino samples are considered together.
The combined sensitivities for wrong mass hierarchy and wrong $\theta_{23}$ octant rejection are shown in Figure \ref{fig:atmos_beam_sensi}.
The mass hierarchy can be determined at 6.2(3.8)\,$\sigma$ for a true value of \sinsqthetaatmos of 0.6(0.4).
The octant can be determined at 6.2(3.6)\,$\sigma$ for a true value of \sinsqthetaatmos of 0.45(0.55).





\section{Proton decay physics potential}

Proton decay is a fundamental prediction of many grand unified theories.
Current limits for lifetime over branching fraction ($\tau/B$) are of the order $10^{33}$--$10^{34}$ years, depending on the specific channel.
Here we discuss the sensitivity to two decay modes.

\subsection{$p \rightarrow e^+ \pi^0$}

The signal of this event is three rings (or two if the $\pi^0$ decay was very asymmetric or forward-boosted).
Cuts are made against the invariant mass of $\pi^0$ (for 3 ring events only), the invariant mass of the proton, and the total momentum.
Events are split between free-proton enriched ($p_{\textrm{tot}} < 100$\,MeV/$c$) and bound-proton enriched ($100 < p_{\textrm{tot}} < 250$\,MeV/$c$) samples.
Atmospheric neutrino events with a muon below threshold are removed by rejecting events with a Michel electron. Neutrons are also rejected.
More details can be found in \cite{sk_pdecay_epi0}.

\begin{figure}[ht]
  \centering
  \includegraphics[width=0.47\textwidth,trim={0 4.5cm 0 4cm},clip]{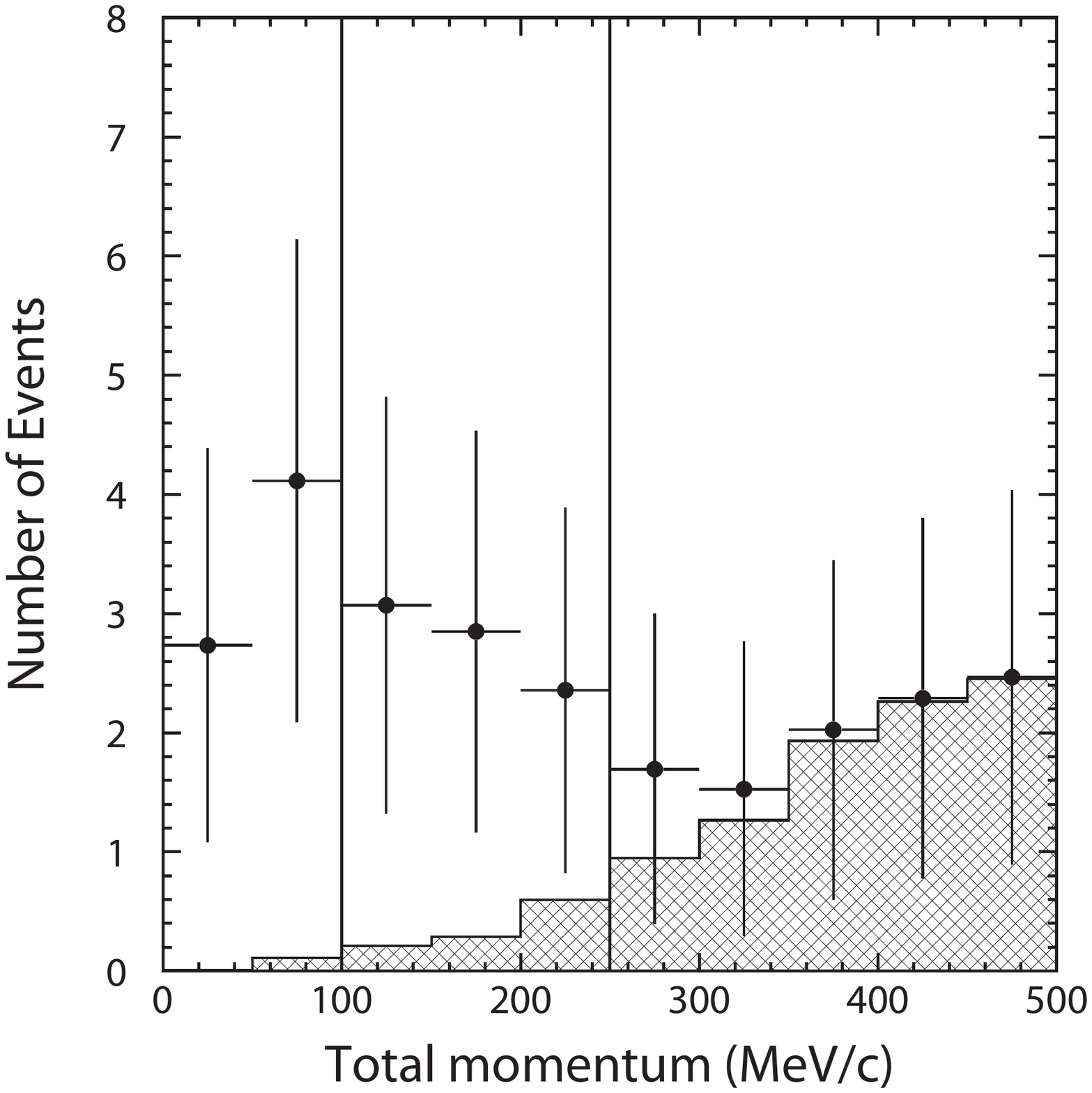}
  \includegraphics[width=0.45\textwidth]{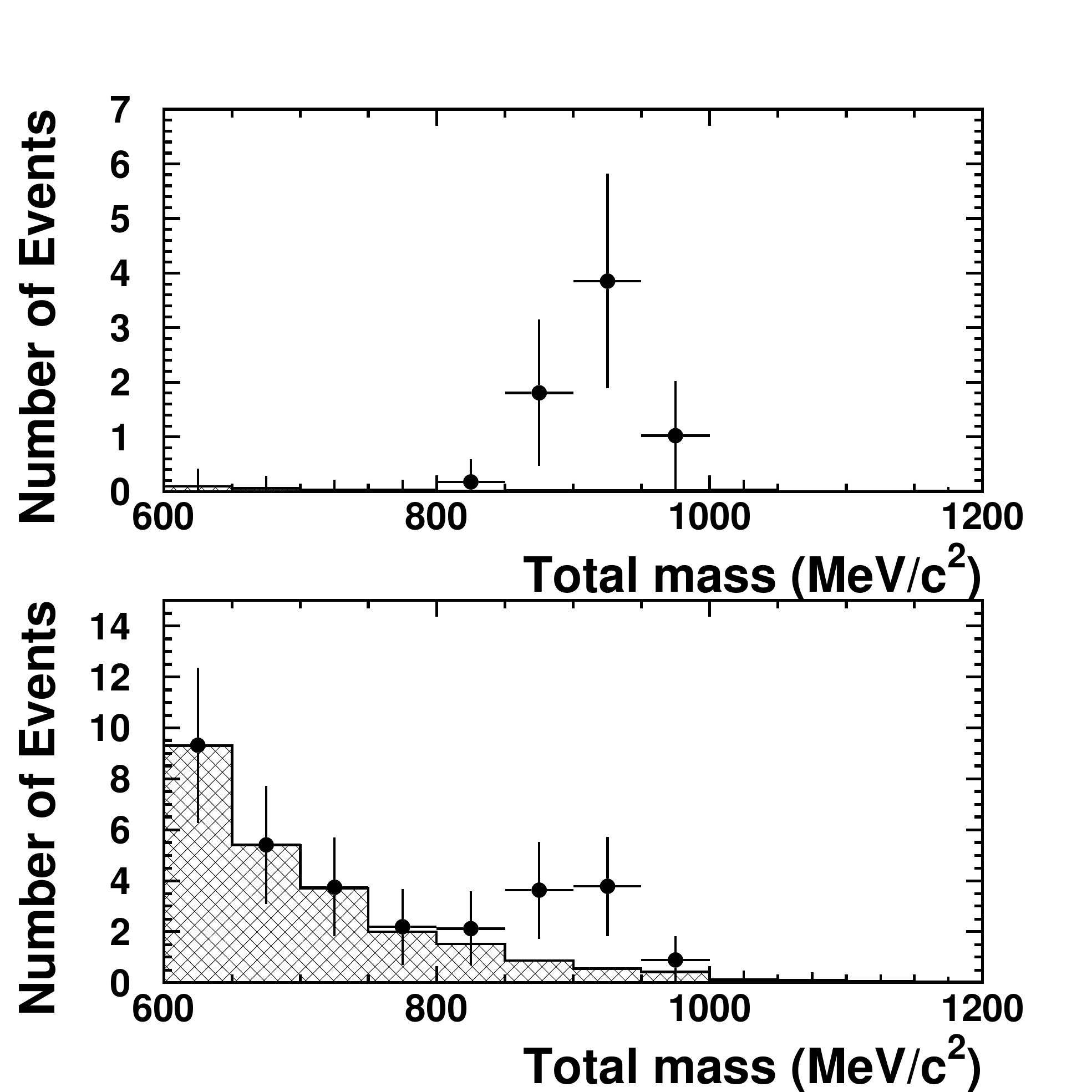}
  \caption{Left: Total momentum distribution of all events passing the $p \rightarrow e^+ \pi^0$ event selection, before the momentum cut.
    Right: Reconstructed invariant mass distribution of all events passing the $p \rightarrow e^+ \pi^0$ event selection, before the invariant mass cut.
    The proton lifetime is assumed to be $1.7 \times 10^{34}$ years, just beyond current Super-K limits.
    The free- and bound-proton enhanced regions are shown by the lines in the left plot, and upper and lower panels of the right plot.
    The atmospheric neutrino background is shown by the hatched region.
  }
  \label{fig:pdecay_epi0_spectra}
\end{figure}

\begin{table}[htb]
  \begin{center}
    \begin{tabular}{|c | c || c | c |}
      \hline
      \multicolumn{2}{|c||}{$p_{\textrm{tot}} < 100$\,MeV/$c$} & \multicolumn{2}{c|}{$100 < p_{\textrm{tot}} < 250$\,MeV/$c$} \\
      $\epsilon_{\textrm{sig}}$ [\%] & Bkg [/Mton$\cdot$yr] & $\epsilon_{\textrm{sig}}$ [\%] & Bkg [/Mton$\cdot$yr] \\
      \hline
      18.7 $\pm$ 1.2 & 0.06 $\pm$ 0.02 & 19.4 $\pm$ 2.9 & 0.62 $\pm$ 0.20 \\
      \hline
    \end{tabular}
    \caption{Signal efficiency and background rates for the $p \rightarrow e^+ \pi^0$ analysis.}
    \label{tab:pdecay_epi0_eff}
  \end{center}
\end{table}

The event distribution, as a function of total momentum and total mass, is shown in Figure \ref{fig:pdecay_epi0_spectra}.
The efficiency and background rates are shown in Table \ref{tab:pdecay_epi0_eff}.
It can be seen that the free-proton enriched sample is almost background free.

The $p \rightarrow e^+ \pi^0$ channel has a 3$\sigma$ discovery potential of $10^{35}$\,yr after 20 years, as shown in the left plot of Figure \ref{fig:pdecay_sensi}.

\subsection{$p \rightarrow \nubar K^+$}

The $K^+$ is produced with momentum 340\,MeV/$c$, significantly below its 749\,MeV/$c$ Cherenkov threshold.
Therefore the analysis needs to identify the decay products: $K^+ \rightarrow \mu^+ \nu$ (64\% branching ratio) and $K^+ \rightarrow \pi^+\pi^0$ (21\% branching ratio).
The events are split into three samples:
1) ``prompt $\gamma$'' searches for a 6.3\,MeV de-excitation $\gamma$ ray\footnote{This occurs when a bound proton decays, and another proton fills the proton hole.} occurring prior to 236\,MeV/$c$ muon;
2) ``$p_\mu$ spectrum'' searches for a 236\,MeV/$c$ muon without associated $\gamma$;
3) ``$\pi^+\pi^0$'' searches for a monochromatic $\pi^0$ with light from background $\pi^+$\footnote{The $\pi^+$ is just above threshold, therefore it is difficult to reconstruct a full ring.}.
More details can be found in \cite{sk_pdecay_nuk}.

\begin{figure}[htb]
  \centering
  \includegraphics[width=0.49\textwidth, trim={0 4.5cm 0 4cm}, clip]{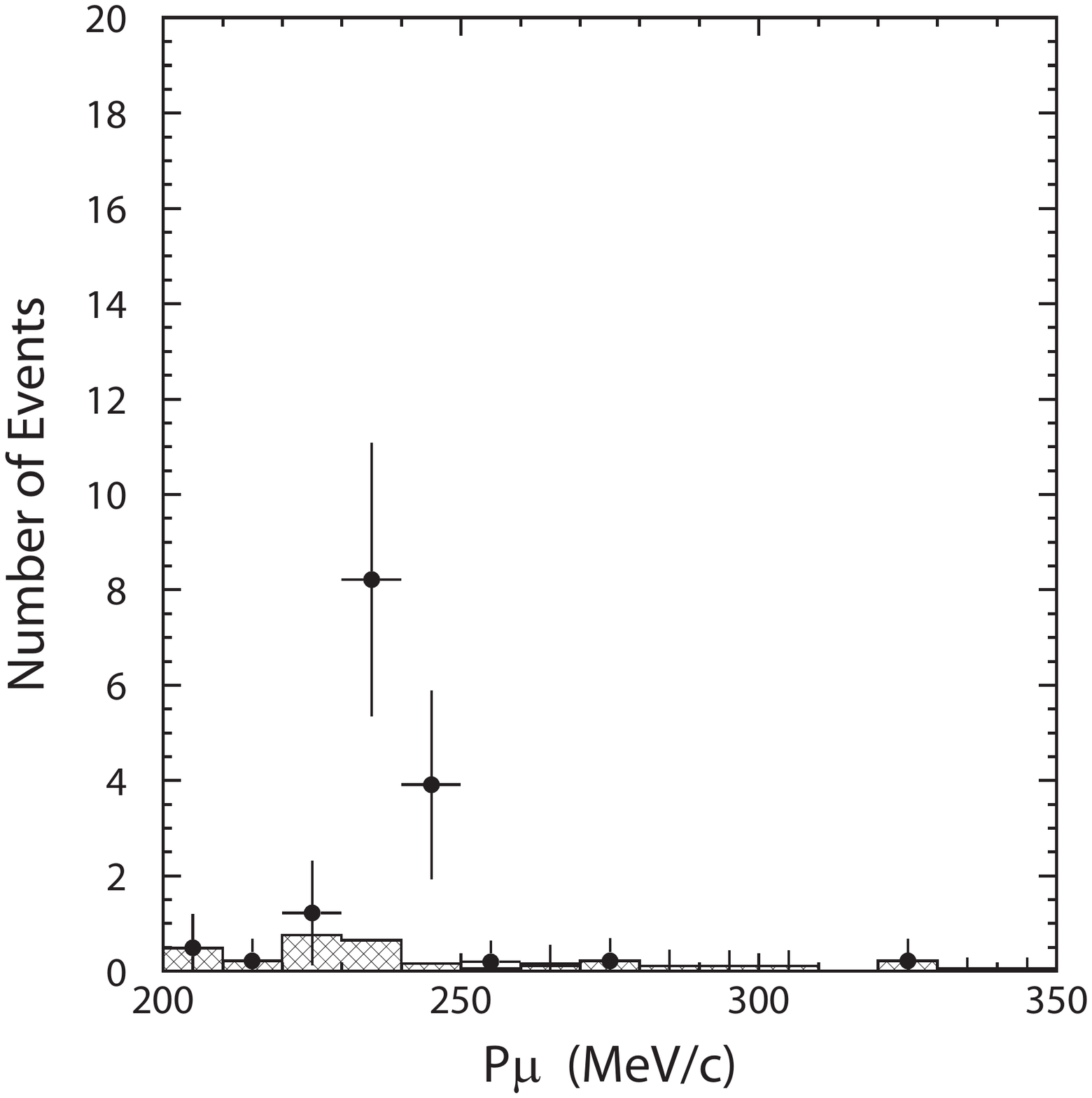}
  \includegraphics[width=0.49\textwidth, trim={0 4.5cm 0 4cm}, clip]{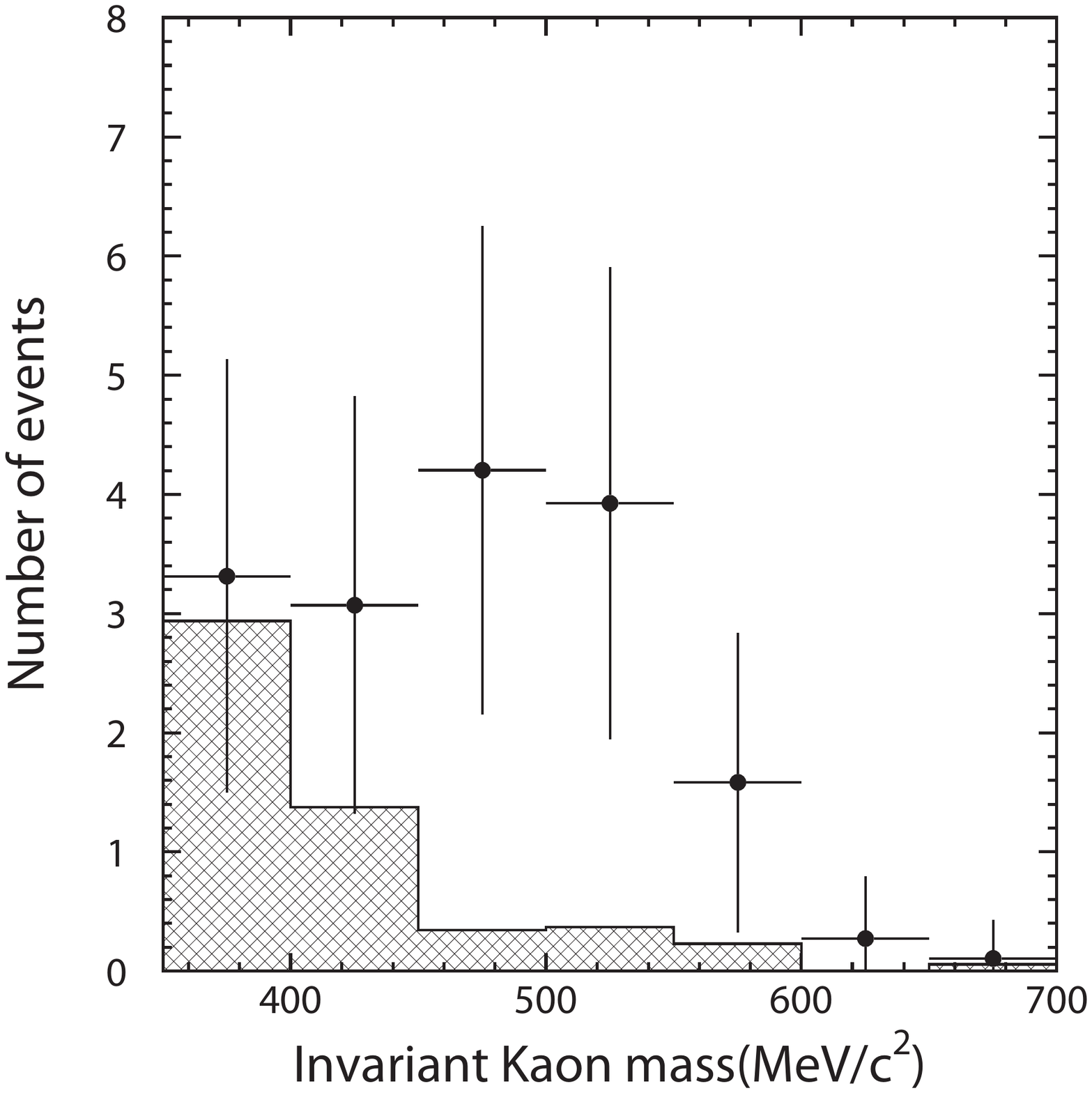}
  \caption{Left: Reconstructed momentum distribution of all events passing the prompt $\gamma$ $p \rightarrow \nubar K^+$ event selection.
    Right: Reconstructed invariant kaon mass distribution of all events passing the $\pi^+\pi^0$ $p \rightarrow \nubar K^+$ event selection, before the visible energy opposite $\pi^0$ candidate cut.
    The proton lifetime is assumed to be $6.6 \times 10^{33}$ years, just beyond current Super-K limits.
    The atmospheric neutrino background is shown by the hatched region.
  }
  \label{fig:pdecay_nuk_spectra}
\end{figure}

Event distributions for the prompt $\gamma$ and $\pi^+\pi^0$ samples are shown in Figure \ref{fig:pdecay_nuk_spectra} as a function of muon momentum and invariant kaon mass respectively.
The efficiency and background rates are shown in Table \ref{tab:pdecay_nuk_eff}.

\begin{figure}[htb]
  \centering
  \includegraphics[width=0.49\textwidth,trim={5cm 3.5cm 5cm 4cm},clip]{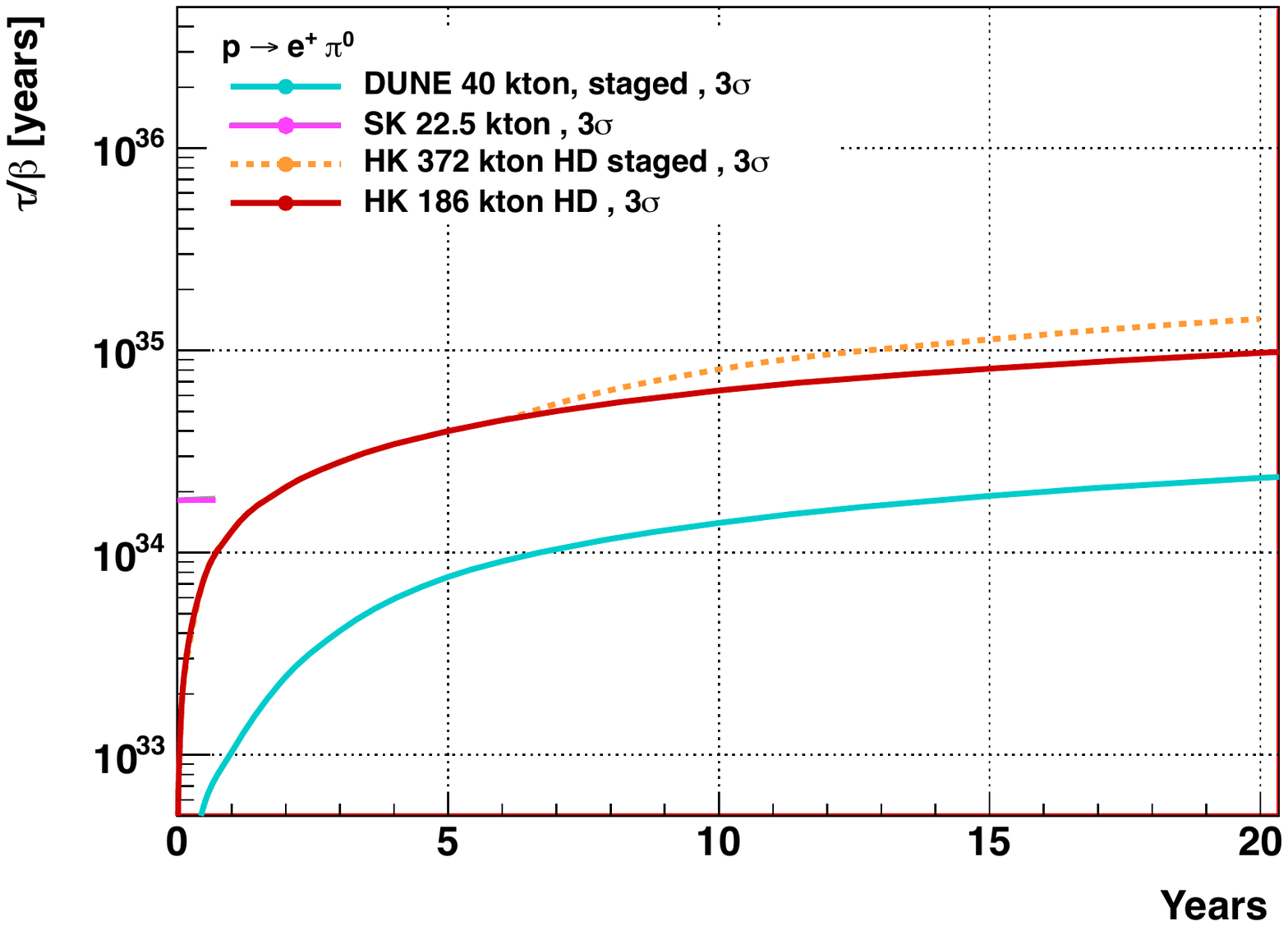}
  \includegraphics[width=0.49\textwidth,trim={5cm 3.5cm 5cm 4cm},clip]{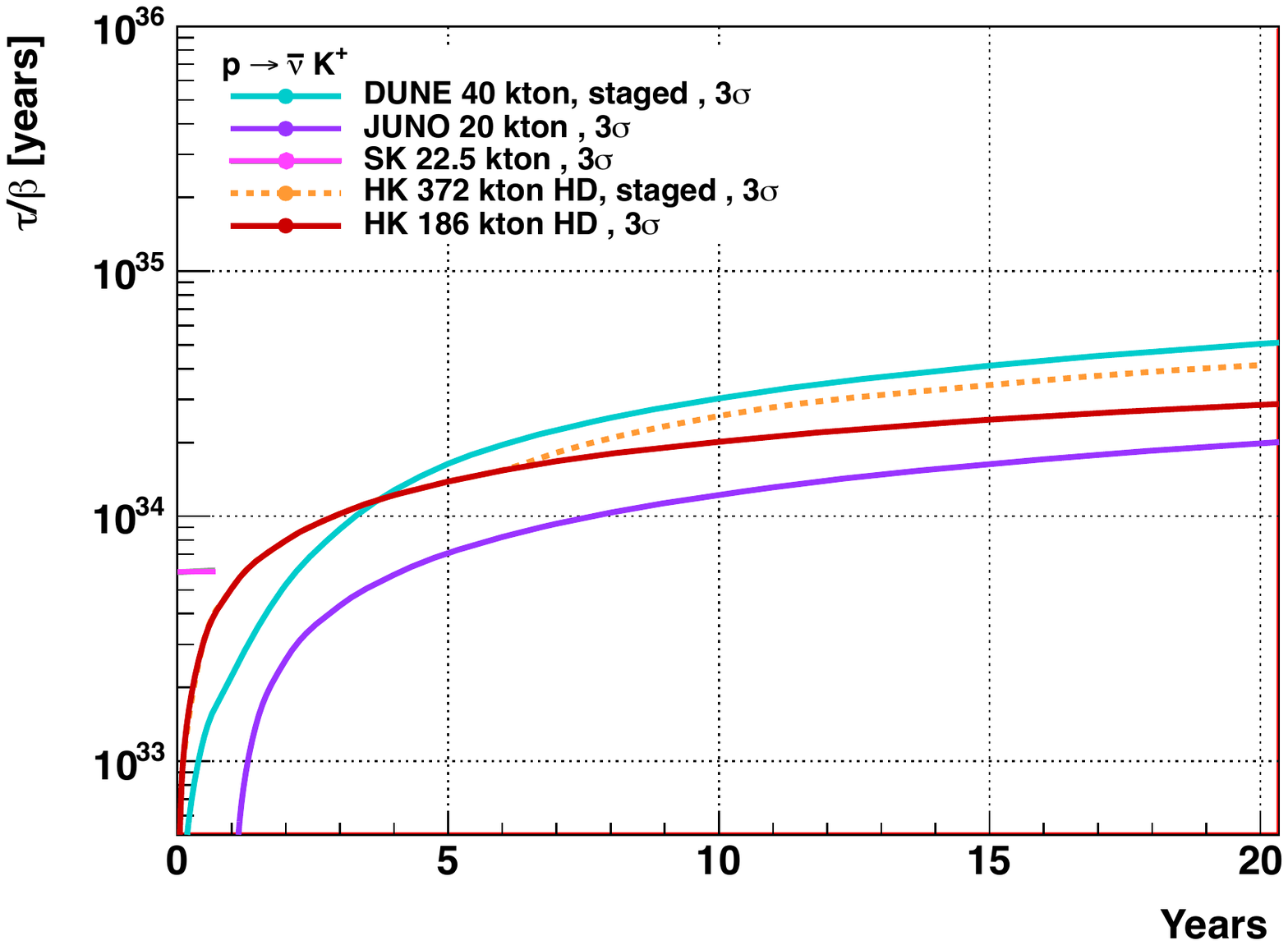}
  \caption{3$\sigma$ discovery potential as a function of time for $p \rightarrow e^+ \pi^0$ (left) and $p \rightarrow \nubar K^+$ (right).
    HK staged refers to a second \hk tank coming online 6 years after the first.
    The DUNE sensitivity is taken from \cite{dune_pdecay} and the Super-K point corresponds to the expectation in 2026 (after 23 years of data).}
  \label{fig:pdecay_sensi}
\end{figure}

The $p \rightarrow \nubar K^+$ channel has a 3$\sigma$ discovery potential of $3 \times 10^{34}$\,yr after 20 years, as shown in the right plot of Figure \ref{fig:pdecay_sensi}.

\begin{table}[htb]
  \begin{center}
    \begin{tabular}{|c | c || c | c || c | c |}
      \hline
      \multicolumn{2}{|c||}{Prompt $\gamma$} & \multicolumn{2}{c||}{$p_\mu$ spectrum} & \multicolumn{2}{c|}{$\pi^+\pi^0$} \\
      $\epsilon_{\textrm{sig}}$ [\%] & Bkg [/Mton$\cdot$yr] & $\epsilon_{\textrm{sig}}$ [\%] & Bkg [/Mton$\cdot$yr] & $\epsilon_{\textrm{sig}}$ [\%] & Bkg [/Mton$\cdot$yr]\\
      \hline
      12.7 $\pm$ 2.4 & 0.9 $\pm$ 0.2 & 31.0 & 1916.0 & 10.8 $\pm$ 1.1 & 0.7 $\pm$ 1.1 \\
      \hline
    \end{tabular}
    \caption{Signal efficiency and background rates for the $p \rightarrow \nubar K^+$ analysis.}
    \label{tab:pdecay_nuk_eff}
  \end{center}
\end{table}

\section{Conclusion}
\hk is the next-generation water Cherenkov experiment to be run in Japan, with data-taking expected to commence $\sim$2027.
In addition to providing CP violation discovery across a wide range of true \dcp values, it has a broad physics program including
mass hierarchy determination within 10 years,
and extending the proton decay lifetime limits by around a factor 10 in some channels.


\end{document}